\documentclass[11pt, a4paper]{article}

\usepackage{amsmath, bm,amssymb}
\usepackage{graphicx, xcolor,float}

\renewcommand\appendix{\par
  \setcounter{section}{0}
  \setcounter{subsection}{0}
  \setcounter{figure}{0}
  \setcounter{table}{0}
  \renewcommand\thesection{Appendix \Alph{section}}
  \renewcommand\thefigure{\Alph{section}\arabic{figure}}
  \renewcommand\thetable{\Alph{section}\arabic{table}}
}
\usepackage{natbib} 
\usepackage[top=1.5 cm, right=1.5cm, left=1.5cm, bottom=2.cm]{geometry}

\title{Comparing robustness properties of optimal designs under standard and compound criteria}

\author{Md.\ Shaddam\ Hossain $\text{Bagmar}^{*a}$, Wasimul $\text{Bari}^{b}$, and A.\ H.\ M.\ Mahbub $\text{Latif}^{c}$\\ \textsuperscript{a}Institute of Statistical Research and Training (ISRT), University of Dhaka, Dhaka-1000, Bangladesh\\ email: shbagmar@isrt.ac.bd\\ \textsuperscript{b}Department of Statistics, University of Dhaka, Dhaka-1000, Bangladesh\\ \textsuperscript{c}Institute of Statistical Research and Training (ISRT), University of Dhaka, Dhaka-1000, Bangladesh and \\ Center for Clinical Epidemiology, St. Luke's International University, 3-6-2 Tsukiji, Chuo-ku,\\ Tokyo 104-0045, Japan} 
\date{}
\begin{document}
\maketitle
\bibliographystyle{apalike}

\begin{abstract}
Standard optimality criteria (e.g.\ {\sl A}-optimality, {\sl D}-optimality criterion, etc.) have been commonly used for obtaining optimal designs. For a given statistical model, standard criteria assume the error variance is known at the design stage. However, in practice the error variance is estimated to make inference about the model parameters. Modified  criteria are defined as a function of the standard criteria and  the corresponding error degrees of freedom, which may lead to extreme optimal design. Compound criteria are defined as the function of different modified criteria and corresponding  user specified weights. Standard, modified, and compound criteria based optimal designs are obtained for $3^3$ factorial design. Robustness properties of the optimal designs are also compared.
\end{abstract}


\section{Introduction}

Statistical design of experiment deals with assigning the treatment combinations of interest to the available experimental units. For a given research question, a number of experimental designs can be considered and the optimal design is the one that ensures efficient estimators of the model parameters. The optimal design helps to make valid conclusion of the experiment. To obtain optimal design, competing experimental designs are compared with respect to a design criterion, which is often defined as a function of the information matrix corresponding  to the statistical model intended to consider for the analysis. 

The commonly used design criteria (such as  {\sl D}-optimality, {\sl A}-optimality, etc.) are known as the standard criteria, which have been widely used in optimal design theory since late  1950's \cite{kiefer1959optimum}.  Considering  the error variance as known at the design stage, the standard criteria are defined as  functions of the information matrix corresponding to the associated statistical model. However, in practice the error variance is estimated using the data obtained from the experiment and the estimated error variance is then used to make inference about the parameters of interest. There is no guarantee that the data obtained from the experiments based on the standard criteria based  optimal designs would provide a reliable estimator of the error variance. Inference based on unreliable estimator of error variance may lead to incorrect conclusion of the experiment. If the error variance is estimated, the properties of the inferences depend on the number of degrees of freedom (df) of the estimator. 
As an extension of the standard optimality criteria, modified optimality criteria are introduced  to accommodate the fact that the error variance is unknown at the design stage. 

Modified optimality criteria are defined as a function of the standard optimality criteria and the quantiles of the appropriate {\sl F}-distributions that are related to test the hypotheses of interest. Modified optimality criteria based optimal design could be very extreme in the sense that it may not allow any lack-of-fit checks, for example. So both the standard and modified optimality criteria have their limitations in not considering estimation of error variance at the design stage and leading to extreme designs, respectively. As a compromise, Gilmour and Trinca \cite{gilmour2012optimum} introduced compound optimality criteria, which are defined as a function of the efficiency of the design with respect to the corresponding standard and modified criteria based optimal designs.

Exact optimal designs depend mainly on four conditions:  the number of experimental runs, research questions under investigation, the  statistical model intended to use for the analysis, and the optimality criterion. Optimal designs obtained for a specific setup (a combination of four conditions) may not be optimal anymore under any violation of one of the underlying conditions. However, in practice the underlying conditions may violate, e.g.\ some observations could be missing,  different models need to be fitted for model selection, etc. 
Therefore, it is of interest to obtain designs that  
provide nearly optimal results even if the underlying conditions are  violated to some extent. Such nearly optimal designs are defined as robust design and  in this paper, robustness properties of a design are quantified against three cases:  missing observations, different model assumptions, and change of design criteria. Considering the  general equivalence theory for minimax optimality criterion, Herzberg and
Andrews \cite{herzberg1976some} 
examined the robustness of polynomial regression models against missing observations. Latif et al. \cite{latif2009robustness}, 
and Ahmad and Gilmour \cite{ahmad2010robustness} 
 also discussed the  robustness against missing  observations in selecting efficient microarray designs and  for subset response surface designs, respectively. To account the model uncertainty, Goos et al. \cite{goos2005model} 
proposed a method to reduce the dependence on the assumed model \cite[see also][]{jones1978design}. For polynomial models with uncorrelated errors, Wong \cite{wong1994comparing} 
studied the robustness of designs under the assumptions  against an incorrect order of the polynomial and  against the change of optimality criteria. 
On the other hand for polynomial models with autocorrelated errors, Moerbeek \cite{moerbeek2005robustness} 
studied the robustness properties of optimal designs against different model assumptions. 
The advisability of comparing designs on the basis of  different criteria of goodness was discussed by Kiefer \cite{kiefer1975optimal}. 
Gilmour and Trinca \cite{gilmour2012optimum} 
studied the robustness properties against  change in criteria considering standard, modified, and compound optimal designs. In this paper, an attempt has been made to examine the robustness properties of the optimal designs against missing observations, change of statistical  model, and change in criteria for different optimal criteria.

In Section \ref{ODcriteria}, standard, modified, and compound  criteria are reviewed.  In Section \ref{robSec}, different robustness measures are discussed, which are robustness  against missing observations, against different model assumptions and against change in criteria. In Section \ref{Result}, robustness properties of standard, modified, and compound criteria based optimal designs are compared.

\section{Optimal design criteria}
\label{ODcriteria}
Consider an hypothetical experiment with $n$ homogeneous experimental units that  are randomly assigned to $t$ treatment combinations $(x_1, \ldots, x_t)$, where $\mathcal{X}=\{x_c\in [-1,1],\;c=1,\ldots, t\}$ is the design space and assume that  at least two experimental units are assigned to one of the treatment combinations. Let $\bm{x}_i=(x_{i_1},\ldots, x_{i_t})$ be the treatment combination assigned to the $i{th}$ experimental unit $(i=1,\ldots, n)$ and  $y_{i}$ be the corresponding response for which the following linear model is assumed
\begin{align}\label{mod1}
y_i&=\sum_{j=1}^p f_j(\bm{x}_{i})\beta_j + \epsilon_i, 
\end{align}
where the $jth$ regression function corresponding to the $ith$ experimental unit $f_j(\bm{x}_{i})$ is either a main effect or  an interaction  of two or more treatment combinations, $\beta_j$ is the regression parameter corresponding to $f_j(\bm{x})$, and the corresponding random error term $\epsilon_i$ is assumed to be independent and identically distributed as normal with mean zero and a constant variance $\sigma^2$. To incorporate intercept in the model, we assume $f_1(\bm{x}_i)=1,\;\forall i$. The model \eqref{mod1} can be expressed in matrix notation as
\begin{align}\label{modMat}
\bm{y}&=\bm{X}\bm{\beta} + \bm{\epsilon},
\end{align}
where $\bm{y}=(y_1, \ldots, y_n)'$, $\bm{X}=(\bm{f}_{1}, \ldots, \bm{f}_{p})$ is the design matrix of order $n\times p$ with $jth$ $(j=1,\ldots, p)$ regression function $\bm{f}_j=(f_j(\bm{x}_1), \ldots, f_j(\bm{x}_n))'$, $\bm{\beta}=(\beta_1, \ldots, \beta_p)'$, and $\bm{\epsilon}=(\epsilon_1, \ldots, \epsilon_n)'$. The maximum likelihood (ML) estimators of $\bm{\beta}$ are the solution of the $p$ system of linear equations
$(\bm{X}'\bm{X})\hat{\bm{\beta}} = \bm{X}'\bm{y}$,
where $\bm{X}'\bm{X}$ is the corresponding information matrix. The expression of the variance-covariance matrix of  $\bm{\hat{\beta}}$,  
$V(\hat{\bm{\beta}}) = (\bm{X}'\bm{X})^{-1} \sigma^{2}$,
is a function of both the information matrix $\bm{X'X}$  and the error variance $\sigma^2$. Note that information matrix is a function of experimental conditions only and in practice, error variance $\sigma^2$ is estimated by the residual mean squares, i.e.\  $\hat{\sigma}^2=(\bm{y}-\bm{X}\bm{\hat{\beta}})'(\bm{y}-\bm{X}\bm{\hat{\beta}})/(n-p-1)$, which is a function of the response and experimental conditions.

Estimating the model parameters with smaller variance ensure making correct conclusions from the experiment. 
For a given number of experimental runs $n$ (say), a number of different combinations of experimental conditions under investigation can be considered. Optimal design is the selection of $n$  conditions that corresponds to the smallest variance of the estimators $\bm{\hat{\beta}}$. A number of design criteria are in the literature that are considered for obtaining optimal designs. Some of those design criteria are briefly discussed in the following sections. 

\subsection{Standard criteria}
Standard criteria,  which are the functions of the type ${R}^p\to{R}$, 
are defined to compare the competing designs in terms of the corresponding information matrices. Among the standard criteria, the most commonly used \textit{D-optimality} criterion is defined for the design with design matrix $\bm{X}$ as the determinant of the corresponding information matrix $\bm{X}'\bm{X}$ as
\begin{equation}
\label{SD}
\phi_{\text{\tiny \sl D}}(\bm{X}) =  \left|\bm{X}'\bm{X}\right|,
\end{equation}
and a design $\xi^\star_{\text{\tiny \sl D}}$ is called the {\sl D}-optimal design if
\begin{equation}
\label{OptMax}
\xi_{\text{\tiny \sl D}}^\star= \operatorname*{arg\; max}_{\bm{X} \in\mathcal{X}}\; \left|\bm{X}'\bm{X}\right| = \operatorname*{arg\; max}_{\bm{X} \in\mathcal{X}}\; \phi_{\text{\tiny \sl D}}(\bm{X}),
\end{equation}
where $\mathcal{X}\in R^t$ is the design space of the experiment. The {\sl D}-optimal design corresponds to the smallest confidence region of the  estimators $\bm{\hat{\beta}}$, which is ensured by maximizing the determinant of the information matrix.

Another important standard criteria is \textit{A-optimality} criterion, which corresponds to minimizing the average variance of the estimators of the model parameters $\bm{\beta}$. Thus,  {\sl A}-optimality criterion is defined for a design with design matrix $\bm{X}$ as  the reciprocal of the average variance
\begin{equation}
\label{Acriterion}
\phi_{A}(\bm{X})= (tr\{W(\bm{X}'\bm{X})^{-1}\})^{-1},
\end{equation}
where $W$ is a diagonal matrix of order $p$, which could be used as the subjective weights to different effects considered in the model  \cite{atkinson1993optimum}. If all the effects are of equal interest, then $W=I_p$ can be considered. 
A design  $\xi_{\text{\tiny \sl A}}^\star$ is called A-optimal design if
\begin{equation}
\label{OptMin}
\xi_{\text{\tiny \sl A}}^\star= \operatorname*{arg\; max}_{\bm{X} \in\mathcal{X}}\; (tr\{W(\bm{X}'\bm{X})^{-1}\})^{-1} = \operatorname*{arg\; max}_{\bm{X} \in\mathcal{X}}\; \phi_{\text{\tiny \sl A}}(\bm{X}).
\end{equation}
Note that {\sl D}- and {\sl A}-optimality criteria are defined as the functions of the information matrix only. 

\subsection{Modified criteria}
Standard criteria are defined under the assumption that the error variance $\sigma^{2}$ is known  at the design stage. Therefore,  the standard criteria based optimal designs do not depend on the estimate of the error variance. However, it   plays an important role in making inference about the parameters of the model. Inefficient estimators of the error variance may lead to incorrect conclusions even if the experiment is conducted with the optimal design. Among the two  estimators of error variance, the pure error df based estimator is more reliable compared to the corresponding estimator mean square error  \citep*{Draper1998}. Thus, the efficiency of the  error variance estimators depends on the size of the corresponding pure error df. To incorporate the effect of  error variance estimator in defining design criterion, modified criterion are defined as a function of pure error df  and the corresponding standard criterion. 

The modified {\sl D}-optimality criterion, which is called {\sl DP}-optimality criterion, is defined for a design with design matrix $\bm{X}$ as
\begin{equation}
\label{DP}
\phi_{\text{\tiny \sl DP}}(\bm{X}, \alpha, d) = \frac{\left|\bm{X}'\bm{X}\right|}{(F_{p, d, (1-\alpha)})^{p}} = \frac{\phi_{\text{\tiny \sl D}}(\bm{X})}{(F_{p, d, (1-\alpha)})^{p}},
\end{equation}
where $\phi_D(\cdot)$ is the standard {\sl D}-optimality criterion, $p$ is the number of parameters in the model, $d$ is the number of pure error df, and $F_{p, d, (1-\alpha)}$ is the $(1-\alpha)$-quantile of the $F$-distribution with $p$ and $d$ df. For a given design, the pure error df can be calculated from the number of times each treatment combination replicated in it. The {\sl DP}-optimal design $\xi_\text{\tiny \sl DP}^\star$ corresponds to the maximum of the {\sl DP}-optimality criterion.

In the same line, the  modified {\sl A}-optimality criterion, which is called  {\sl AP}-optimality criterion, is defined for a design with design matrix $\bm{X}$ as 
\begin{align}
\label{AP}
\phi_{\text{\tiny \sl AP}}(\bm{X}, \alpha, d) = (F_{1, d, 1-\alpha}\; tr\{W(\bm{X}'\bm{X})^{-1}\})^{-1} = (F_{1, d, 1-\alpha})^{-1}\; \phi_{A}(\bm{X}),
\end{align}
where $\phi_A(\cdot)$ is the standard {\sl A}-optimality criterion and $W$ is defined in equation ($\ref{Acriterion}$).  The {\sl AP}-optimal design $\xi_\text{\tiny \sl AP}^\star$ corresponds to the maximum of the {\sl AP}-optimality criterion \eqref{AP}. 
Because of incorporating an $F$-statistic, the modified criteria based optimal designs could be extreme designs and such designs  may not be very useful in practice, e.g.\  in examining the lack-of-fit of the assumed model. Thus, the standard and modified criteria have their limitations and to overcome these limitations,  a combination of standard and modified criteria  is considered as a design criterion. Such criteria are known as compound criteria, which are briefly described in the following sections. 

\subsection{Compound criteria}
Gilmour and Trinca \cite{gilmour2012optimum} strongly argued for  a criterion that would be a combination of different criteria instead of an individual standard or modified criterion. 
To define a general criterion, the analysis of experiment can be  classified into different categories with the expectation that the objective of an experiment will fall in one or more of these categories. For  this purpose, the following efficiencies are defined for the design matrix $\bm{X}$ which has $d$ df  for pure error:

\begin{itemize}
\item[(i)] The {\sl DP}-optimality criterion is used to obtain the optimal design if a global $F$-test will be used in the analysis. The efficiency with respect to the {\sl DP}-optimal design ({\sl DP}-\textit{efficiency}) is defined as
\begin{align*}
E_{\text{\tiny \sl DP}}(\bm{X}) = \bigg[\frac{\phi_{\text{\tiny \sl DP}}(\bm{X}, \alpha, d)}{\phi_{\text{\tiny \sl DP}}(\bm{X}_{\text{\tiny \sl DP}}, \alpha, d_{\text{\tiny \sl D}})}\bigg]^{1/p},
\end{align*}
where $\bm{X}_{\text{\tiny \sl DP}}$ is the design matrix corresponding to the {\sl DP}-optimal design $\xi_{\text{\tiny \sl DP}}^\star$ that corresponds to the maximum of $\phi_{\text{\tiny \sl DP}}$ and $d_\text{\tiny \sl D}$ is the corresponding pure error df. 

\item[(ii)] The weighted {\sl AP}-optimality criterion is used to test individual treatment parameters ($t$-test) and the corresponding \textit{Weighted AP-efficiency} is defined as
\begin{equation*}
E_{\text{\tiny \sl AP}}(\bm{X}) = \frac{\phi_{\text{\tiny \sl AP}}(\bm{X}, \alpha, d)}{\phi_{\text{\tiny \sl AP}}(\bm{X}_{\text{\tiny \sl DP}}, \alpha, d_\text{\sl\tiny A})}, 
\end{equation*}
where $\bm{X}_{\text{\tiny \sl AP}}$ is the design  matrix of the weighted {\sl AP}-optimal design $\xi_{\text{\tiny \sl AP}}^\star$ that corresponds to the maximum of $\phi_{\text{\tiny \sl AP}}$ and  $d_{A}$ is the corresponding  pure error df. 

\item[(iii)]\textit{Degrees-of-freedom efficiency} ({\sl DF}-efficiency) is used for checking the lack of fit of the assumed treatment model and is defined  as
\begin{align}
E_{\text{\tiny \sl DF}}(\bm{X}) &=\frac{(n - d)}{n}.
\end{align}
The {\sl DF}-efficiency is the proportion of experimental resource which is used to estimate the effect of treatments \citep*{daniel1976applications}. As the pure error df  ($d$) decreases, {\sl DF}-efficiency increases. This could be helpful  to overcome the shortcomings of modified criteria to ensure sufficient number of df  for lack-of-fit checking.
\end{itemize}
According to Gilmour and Trinca \cite{gilmour2012optimum}, the compound criteria is defined as 
\begin{align}\label{COMcri}
\phi_C(\bm{X}) &= \big[E_{\text{\tiny \sl DP}}(\bm{X})\big]^{\kappa_{1}}\;\times\; \big[E_{\text{\tiny \sl AP}}(\bm{X})\big]^{\kappa_{2}}\;\times\; \big[E_{\text{\tiny \sl DF}}(\bm{X})\big]^{\kappa_{3}},
\end{align}
where $\kappa_1$, $\kappa_2$, and $\kappa_3$ are non-negative weights corresponding to the {\sl DP}-, {\sl AP}-, and {\sl DF}-efficiency, respectively,  such that $\sum_{l=1}^3\kappa_l=1$. A large value of the weight indicates the importance of the corresponding efficiency. Different compound designs can be considered for different combination of values of $\kappa$'s. 

\section{Robustness}\label{robSec}
Robustness property of optimal design is defined as its ability to perform effectively even when the associated underlying assumptions are violated. In this section, robustness properties of optimal design are defined in three different contexts: (a) under missing observations, (b) under different model assumptions, and (c) against the change in optimality criterion. 

\subsection{Robustness under missing observations}
For a linear model of the type \eqref{modMat},  criteria for robustness under missing observations can be defined in terms of  the generalized variance $(\bm{X}'\bm{D}^{2}\bm{X})^{-1}$,  where $\bm{X}$ is the design  matrix of order $n\times p$ and $\bm{D}^{2}$ is a $n$-dimensional diagonal matrix with diagonal elements  are either zero or one \cite{herzberg1976some}. The number of zeros in the diagonal elements of $\bm{D}^2$ corresponds to the number of missing observations in the data, i.e.\ $\bm{D}^2=\bm{I}_n$ if there is no missing observation and for $m_0$ $(1\leq m_0 <n)$ missing observations $m_0$ rows of $\bm{D}^2$ is replaced by $\bm{0}_p'$. Let $\mathcal{D}_{(s)}$ be the set of $n\choose s$  matrices that can be obtained from $\bm{D}^2$ with $s$ zeros and $(n-s)$ ones in the diagonal elements.

The simplest criterion of robustness under missing observations is the \emph{breakdown number} (BdN), which is defined as  the minimum number of missing observations for which the effect of interest is no longer estimable, i.e.\ $\left|\bm{X}'\bm{D}^2\bm{X}\right|=0$. Latif et al. \cite{latif2009robustness} discussed the breakdown number in the context of microarray experiments. The breakdown number of a design with design  matrix $\bm{X}$ is defined as  
\[
\text{BdN}(\bm{X})=\operatorname*{arg\, min}_{s\in\{1,  \dots, n-1\}}
\big\{\forall\;\bm{D}^2\in\mathcal{D}_{(s)}: \left|\bm{X}'\bm{D}^2\bm{X}\right|=0\big\}.\]
A large  value of BdN leads to more robust design.                
Similar to breakdown number, \emph{probability of breakdown} (BdP) can also be used as a robustness criterion  to quantify the robustness under missing observations. For a design with the design  matrix $\bm{X}$, the probability of breakdown is  defined  as \[\text{BdP}(\bm{X})=P\big(\left|\bm{X}'\bm{D}^{2}\bm{X}\right|=0\big).\] A small value of the probability of breakdown leads to more robust design. The probability of breakdown is estimated numerically by generating a large number of $\bm{D}^2$ matrices with a pre-specified probability of missing observations $p_m\in(0, 1)$ (say), and the proportion of $\left|\bm{X}'\bm{D}^2\bm{X}\right|=0$ is used as the estimate of the probability of breakdown.

Andrews and Herzberg \cite{andrews1979robustness} suggested another criterion for robustness under missing observations, which is based on the estimated variance of the predicted response $V(\hat{\bm{y}})=\bm{H}\sigma^2$, where $\bm{H}=\bm{X}(\bm{X}'\bm{X})^{-1}\bm{X}'$ is known as the hat matrix in regression model literature. The robustness criterion is defined as $\sigma_{v}^{2} = \sum_{i=1}^{n} ( v_{ii} - \bar{v})^{2}/n$, where $v_{ii}=H_{ii}$, the $ith$ diagonal element of the hat matrix $\bm{H}$, and $\bar{v} = \sum_{i=1}^{n}(v_{ii}/n)$. A small value of $\sigma^2_v$ leads to more robust design.

\subsection{Robustness under model assumptions}
Let $\xi_{k, m}^\star$ be the $k$-optimal design corresponding to the the model $M_m$ and $\bm{X}_{k, m}$ be the corresponding model  matrix with $k\in\mathcal{K}$, where $\mathcal{K}$ is the set of optimal design criteria under consideration.  The robustness under model assumption of the design $\xi_{k, m}^\star$ with respect to another model $M_{m'}$, which is nested under $M_m$, 
is formally defined as 
\begin{equation}
\psi_2(k, M_{m}, M_{m'}) = \phi_k(\bm{X}_{k^\star})/\phi_k(\bm{X}_{k, m}), 
\end{equation}
where $\bm{X}_{k^\star}$ is the design  matrix that contains only those factors that are common to both the models $M_m$ and $M_{m'}$. This is an important consideration because most optimal designs are model specific and the true model is usually unknown in practice. Robustness under different model assumptions implies how sensitive the optimality criteria under fitting wrong model considering the true model known.

\subsection{Robustness under change of optimality criteria}
For a given model $m$, let $\xi_{k, m}^\star$ be the $k$-optimal design and $\bm{X}_k$ be the corresponding deign matrix, $k\in\mathcal{K}$, where  $\mathcal{K}$ is the set of optimal design criteria under consideration. 
The robustness under different optimality criteria  of the design $k$-optimal design $\xi^\star_{k, m}$ with respect to the  criterion $k'\neq k\in\mathcal{K}$ can be defined as
\begin{equation}
\label{psi3}
  \psi_3(\bm{X}_k, \bm{X}_{k'})=\phi_{k'}(\bm{X}_{k})/\phi_{k'}(\bm{X}_{k'}),
\end{equation}
where $\bm{X}_{k'}$ is the design  matrix corresponding to the $k'$-optimal design $\xi_{k', m}^\star$ and $\psi_3$ takes the value in the interval $(0, 1]$.

\section{Robustness properties of optimal designs}\label{Result}
In this section, standard, modified, and compound criteria based optimal designs  are compared on the basis of the three robustness properties  defined in Section~\ref{robSec}.  Among the standard and modified criteria, the {\sl D}-, {\sl A}-,  {\sl DP}-, and {\sl AP}-optimality criteria are considered  for the comparison.  Compound criteria can be defined for different set of $\bm{\kappa}=(\kappa_1, \kappa_2, \kappa_3)$ values in the expression of $\phi_C$  defined in \eqref{COMcri}. In this paper, two compound criteria {\sl C1} and {\sl C2} are considered that correspond to the $\bm{\kappa}$ values  (.8. 0, .2) and (0, .8, .2), respectively in the expression of $\phi_C$. 

\subsection{Optimal designs}\label{odes}
 The following four models are considered to obtain standard, modified, and compound criteria based optimal designs
\begin{align}
 y_i  &= \beta_{0} + \sum_{j=1}^3 \beta_{j} x_{j(i)} + \epsilon_{i} \tag{$M_1$} \label{nmod1} \\ 
 y_i  &= \beta_{0} + \sum_{j=1}^3 \big(\beta_{j} x_{j(i)} + \beta_{jj} x_{j(i)}^2\big) + \epsilon_i \tag{$M_2$} \label{nmod2} \\
y_i  &= \beta_{0} + \sum_{j=1}^3 \beta_{j} x_{j(i)}  +\sum_{j>j'} \beta_{jj'} x_{j(i)} x_{j'(i)} +\epsilon_i, \tag{$M_3$} \label{nmod3} \\
y_i & = \beta_{0} + \sum_{j=1}^3 \big(\beta_{j} x_{j(i)}  + \beta_{jj} x_{j(i)}^2\big)  + \sum_{j'>j}^3 \beta_{j'j} x_{j'(i)} x_{j(i)} + \epsilon_i,\tag{$M_4$}
\label{nmod4}
\end{align}
where $x_{j(i)}$ be the level of the factor $x_j\in\{-1, 0, 1\}$ that is randomly assigned to the $ith$ run of the experiment $(i=1,\ldots, n)$, $\beta$'s are regression parameters, and random error term $\epsilon$ is assumed to be independent and normally distributed with mean 0 and a constant variance $\sigma^2$. 
The models \eqref{nmod1}--\eqref{nmod4} contain different combinations of linear, quadratic, and interaction terms.  The  model \eqref{nmod1}  is the simplest one that contains only the linear terms,  the model \eqref{nmod2} contains the linear and quadratic terms, the model \eqref{nmod3} contains  the linear  and interaction terms, and the model \eqref{nmod4} contains all the linear, quadratic, and interaction terms of the factors $x_1$, $x_2$, and $x_3$. So the model \eqref{nmod1} is nested under the other three models \eqref{nmod2}--\eqref{nmod4}, and the models \eqref{nmod2}  and \eqref{nmod3} are nested under the model \eqref{nmod4} only.  The standard, modified, and compound criteria based exact optimal designs  for the models \eqref{nmod1}--\eqref{nmod4}, each with $n=16$ runs, are obtained using standard exchange algorithm \citep*{atkinson2007optimum}   and are presented in the Tables \ref{MIS1}--\ref{MIS4}, where the selected treatment combinations and the number of times it repeated are reported for all the optimal designs. 

Table~\ref{MIS1} shows the optimal designs for the model \eqref{nmod1}. The {\sl D}- and {\sl A}-optimal designs  ($\xi^\star_{\text{\tiny \sl D}, \text{\sl 1}}$ and $\xi^\star_{\text{\tiny \sl A}, \text{\sl 1}}$) consist of the same eight treatment combinations, where each of the eight treatment combinations is repeated  twice for the design $\xi^\star_{\text{\tiny \sl D}, \text{\tiny 1}}$ and  for  $\xi^\star_{\text{\tiny \sl A}, \text{\tiny 1}}$, four of the treatment combinations repeated three times and the other four repeated one time each. So, the  pure error df  (pedf) is 8 for both the designs. The same four treatment combinations, each repeated four times,  are selected for the {\sl DP}- and {\sl AP}-optimal designs ($\xi^\star_{\text{\tiny \sl DP}, \text{\tiny 1}}$ and $\xi^\star_{\text{\tiny \sl AP}, \text{\tiny 1}}$) and  the pedf is 12 for both the designs. The {\sl C1}- and {\sl C2}-optimal designs ($\xi^\star_{\text{\tiny \sl C1}, \text{\tiny 1}}$ and $\xi^\star_{\text{\tiny \sl C2}, \text{\tiny 1}}$) consist of 12 and 13 different treatment combinations and the corresponding pedfs are 4 and 3, respectively. 

Table~\ref{MIS2} shows the optimal designs for  the model~\eqref{nmod2}. The pedfs for the {\sl D}- and {\sl A}-optimal designs ($\xi^\star_{\text{\tiny \sl D}, \text{\tiny 2}}$ and $\xi^\star_{\text{\tiny \sl A}, \text{\tiny 2}}$) are 1 and 0, respectively, for {\sl DP}- and {\sl AP}-optimal designs ($\xi^\star_{\text{\tiny \sl DP}, \text{\tiny 2}}$ and $\xi^\star_{\text{\tiny \sl AP}, \text{\tiny 2}}$) are 7 and 6, respectively, and for {\sl C1}- and {\sl C2}-optimal designs ($\xi^\star_{\text{\tiny \sl C1}, \text{\tiny 2}}$ and $\xi^\star_{\text{\tiny \sl C2}, \text{\tiny 2}}$) are 4 and 3, respectively. For the model \eqref{nmod3}, the standard and modified criteria based optimal designs are similar to the {\sl D}-optimal design for the model~\eqref{nmod1}, which is a complete run of a $2^3$ design with each of the three factors has levels $-1$  and $+1$. As expected, the {\sl C1}- and {\sl C2}-optimal designs are different than the standard and modified criteria based optimal designs and the corresponding pure error df  is 4 for both the designs. 
The optimal designs for the model \eqref{nmod4} are presented in Table~\ref{MIS4}, which shows that 
the pure error df  for both the {\sl D}- and {\sl A}-optimal designs ($\xi_{\text{\tiny \sl D}, \text{\tiny 4}}^{\star}$ and $\xi_{\text{\tiny \sl A}, \text{\tiny 4}}^{\star}$) is 0, for the  {\sl DP}- and {\sl AP}-optimal designs ($\xi_{\text{\tiny \sl DP}, \text{\tiny 4}}^{\star}$ and $\xi_{\text{\tiny \sl AP}, \text{\tiny 4}}^{\star}$) are 6 and 5 respectively, and for the  {\sl C1}- and {\sl C2}-optimal designs ($\xi_{\text{\tiny \sl C1}, \text{\tiny 4}}^{\star}$  and $\xi_{\text{\tiny \sl C2}, \text{\tiny 4}}^{\star}$) are 4 and 3, respectively. The pure error df  for different optimal design are shown in Table \ref{tpedf1}.


\begin{table}[H]
\centering
\caption{The pure error df  of the optimal designs for the models \eqref{nmod1}--\eqref{nmod4}}\medskip
\label{tpedf1}
{\begin{tabular}[l]{@{}ccccc}
\hline
Design criteria & \eqref{nmod1}& \eqref{nmod2}& \eqref{nmod3}& \eqref{nmod4} \\
\hline
{\sl D} & 8 & 1 & 8 & 0 \\
{\sl A} & 8 & 0 & 8 & 0 \\
{\sl DP} & 12 & 8 & 8 & 5 \\
{\sl AP} & 12 & 7 & 8 & 4 \\
{\sl C1} & 4 & 4 & 4 & 3 \\
{\sl C1} & 3 & 3 & 4 & 3 \\
\hline
\end{tabular}}
\end{table}

\subsection{Robustness under missing observations}
\label{Miss}

Table~\ref{MISa} shows the estimates  of different measures of robustness under missing observations, namely breakdown probability (BdP), breakdown number (BdN) and $\sigma^{2}_{v}$, for different optimal designs obtained for the models \eqref{nmod1}--\eqref{nmod4} with $n=16$ runs. For calculating breakdown probabilities for each of the  optimal designs described in \S\ref{odes},  the $\bm{D}^2$ matrices are generated 1000 times with a pre-specified probability of missing observations, which are 0.40 for the model \eqref{nmod1} and 0.40 for the other models. The results show that the robustness property under missing observations depends on both the underlying model and the criteria used for quantifying it. Based on the robustness criterion BdN, {\sl A}-, {\sl C1}-, and {\sl C2}-optimal designs are found to be the most robust for the model \eqref{nmod1}, {\sl A}- and {\sl C1}-optimal designs for the model \eqref{nmod2}, all the competing optimal designs except the {\sl A}-optimal design for the model \eqref{nmod3}, and {\sl A}-, {\sl D}-, and {\sl C2}-optimal designs for the model \eqref{nmod4}. On the other hand, based on the criterion BdP the {\sl C2}-optimal design is found to be the most robust for the model \eqref{nmod1}, {\sl D}- and {\sl A}-optimal designs for the model \eqref{nmod2}, {\sl C1}- and {\sl C2}-optimal designs for \eqref{nmod3}, and {\sl A}-optimal design for the model \eqref{nmod4}. The criterion $\sigma^2_v$ is not found very useful in finding the most robust optimal design for the models \eqref{nmod1} and \eqref{nmod3} as it takes the value zero for some of the designs. The estimates of $\sigma^2_v$ show that the {\sl D}-optimal design for the model \eqref{nmod2}, and {\sl D}- and {\sl A}-optimal designs for the the model \eqref{nmod4} are the most robust.

%


\begin{table}[H]
\small
\centering
\caption{Estimated robustness criteria under missing observations, BdP, BdN, and $\sigma_v^2$, for standard, modified, and compound criteria based optimal designs  for the models \eqref{nmod1}--\eqref{nmod4} with $n=16$ runs. For calculating BdP, 0.40 is considered probability of missing observations for the model \eqref{nmod1} and for other models 0.20 is considered.}\medskip
\label{MISa}
\begin{tabular}{ccccccccccccccccc}
\hline
Design & & \multicolumn{3}{c}{(\ref{nmod1})} &&  \multicolumn{3}{c}{(\ref{nmod2})} &&    \multicolumn{3}{c}{(\ref{nmod3})} && \multicolumn{3}{c}{(\ref{nmod4})} \\ 
\cline{3-5} \cline{7-9} \cline{11-13} \cline{15-17}
criteria&&BdP &BdN & $\sigma^{2}_{v}$&& BdP& BdN&$\sigma^{2}_{v}$&& BdP& BdN& $\sigma^{2}_{v}$&& BdP &BdN&$\sigma^{2}_{v}$\\ \cline{1-1} 
\cline{3-5} \cline{7-9} \cline{11-13} \cline{15-17} 
{\sl D}  && 0.012& 7 & 0&&0.004&4  & 0.001&& 0.038&4 &  0&& 0.019 &3  &0.014 \\
{\sl A} && 0.008&8  & 0&&0.004 &5  & 0.003&& 0.038&4 & 0&&  0.017 &3 &0.014  \\ 
{\sl DP} && 0.097&4  & 0&& 0.043&3  & 0.007&& 0.036&2 &0.007  && 0.116 &1 &0.047  \\
{\sl AP}&& 0.098&4  & 0&& 0.045&3 & 0.007&& 0.039&4 &0  && 0.097 &2  &0.033\\
{\sl C1} && 0.004&8 & 0.001&& 0.006&5 & 0.003&& 0.015&4 &0.009 && 0.079 &2 &0.027 \\
{\sl C2}&& 0.003&8 & 0.001&& 0.006 &4 & 0.002&& 0.015&4 &0.009 && 0.099 &3  &0.016 \\ \hline 
\end{tabular}
\end{table}

\subsection{Robustness under different model assumptions}
\label{DMA}

The estimated criteria of robustness under different model assumptions are reported in  Table~\ref{MR1}(a)--\ref{MR1}(c) for the optimal designs obtained in \S\ref{odes}. Table~\ref{MR1}(a)  shows the performance of the optimal designs obtained for the models \eqref{nmod2}--\eqref{nmod4} if they are used for the model \eqref{nmod1}. Similarly, Tables~\ref{MR1}(b) and \ref{MR1}(c) show the performance of the optimal designs obtained for the model \eqref{nmod4}  if they are used for the models \eqref{nmod2} and  \eqref{nmod3}, respectively. The compound criteria based optimal designs are found to be the most robust under different model assumption for all three models \eqref{nmod2}--\eqref{nmod4}, and the modified criteria based optimal designs are found to be the least robust.


\begin{table}[H]
\centering
 \caption{Estimates of robustness under model assumption criterion for standard, modified, and compound criteria based optimal designs  with $n=16$ runs. $(a)$ under the model \eqref{nmod1} when the fitted the models are \eqref{nmod2}--\eqref{nmod4}, $(b)$ under the model \eqref{nmod2} when the fitted model is \eqref{nmod4}, and $(c)$ under the model \eqref{nmod3} when the fitted model is \eqref{nmod4}.
}\medskip
 \label{MR1}
\small
\begin{tabular}{cccccccccccc}
\multicolumn{6}{c}{$(a)$} && \multicolumn{2}{c}{$(b)$} && \multicolumn{2}{c}{$(c)$} \\
\cline{1-6} \cline{8-9} \cline{11-12}
Optimal design &\eqref{nmod2}&&\eqref{nmod3} && \eqref{nmod4} && Optimal design & \eqref{nmod4}  && Optimal design & \eqref{nmod4} \\ 
\cline{1-6} \cline{8-9} \cline{11-12}
$\xi_{\text{\tiny \sl A}, \text{\tiny 1}}^\star$  &  0.704 && 1.000 && 0.755&& $\xi_{\text{\tiny \sl A}, \text{\tiny 2}}^\star$& 0.960&&  $\xi_{\text{\tiny \sl A}, \text{\tiny 3}}^\star$ &0.651 \\ 
$\xi_{\text{\tiny \sl D}, \text{\tiny 1}}^\star$  &0.632 && 1.000 && 0.757&& $\xi_{\text{\tiny \sl D}, \text{\tiny 2}}^\star$  &  0.956&&  $\xi_{\text{\tiny \sl D}, \text{\tiny 3}}^\star$&0.666 \\
$\xi_{\text{\tiny \sl DP}, \text{\tiny 1}}^\star$ &0.597 && 0.893 && 0.575&& $\xi_{\text{\tiny \sl DP}, \text{\tiny 2}}^\star$ &  0.795 &&  $\xi_{\text{\tiny \sl DP}, \text{\tiny 3}}^\star$& 0.510 \\
$\xi_{\text{\tiny \sl AP}, \text{\tiny 1}}^\star$ &0.598 && 0.893 && 0.499&& $\xi_{\text{\tiny \sl AP}, \text{\tiny 2}}^\star$ &  0.638&&  $\xi_{\text{\tiny \sl AP}, \text{\tiny 3}}^\star$& 0.506  \\
$\xi_{\text{\tiny \sl C1}, \text{\tiny 1}}^\star$ & 0.937 && 1.000 && 0.956&& $\xi_{\text{\tiny \sl C1}, \text{\tiny 2}}^\star$&  0.975&&  $\xi_{\text{\tiny \sl C1}, \text{\tiny 3}}^\star$ &   0.946 \\
$\xi_{\text{\tiny \sl C2}, \text{\tiny 1}}^\star$ & 0.955 && 0.996 && 0.977&&  $\xi_{\text{\tiny \sl C2}, \text{\tiny 2}}^\star$&  0.980 &&  $\xi_{\text{\tiny \sl C2}, \text{\tiny 3}}^\star$& 0.965  \\ 
\hline
 \end{tabular}
\end{table}

\subsection{Robustness under change of optimality  criteria}\label{Change}
The standard, modified, and compound criteria based optimal designs, which are obtained for the models \eqref{nmod2}--\eqref{nmod4} and described in \S\ref{odes}, are compared with respect to the robustness under change of optimality criteria $\psi_3$, defined in \eqref{psi3}. Table~\ref{RCC2} shows the efficiencies of optimal designs obtained for the models \eqref{nmod2}--\eqref{nmod4} with respect to different optimality criteria. The {\sl A}-optimal designs are found to be highly  ($\simeq 100\%$) efficient with respect to  {\sl D}-optimality criterion  for the models \eqref{nmod2}--\eqref{nmod4}. However, the modified and compound criteria cannot be evaluated with the {\sl A}-optimal designs for the models \eqref{nmod2} and \eqref{nmod4}, i.e.\ for  $\xi_{\text{\sl \tiny A}, \text{\tiny 2}}^\star$  and $\xi_{\text{\sl \tiny A}, \text{\tiny 4}}^\star$. The {\sl A}-optimal design for the model~\eqref{nmod3} is found to be 100\% efficient with respect to the  standard and modified criteria, and about 80\%  efficient with respect to  the compound criteria. The performance of the {\sl D}-optimal designs are similar to that of the {\sl A}-optimal designs, except for the model \eqref{nmod2} for which  the {\sl D}-optimal designs can be evaluated for the modified and compound criteria. The {\sl DP}- and {\sl AP}-optimal designs are found to be highly efficient with respect to other competing optimality criteria for all the models and overall, the {\sl AP}-optimal designs are more robust under the change of optimality criteria than the {\sl DP}-optimal designs. Similar to the  modified criteria based optimal designs, the compound criteria based optimal designs (i.e.\ {\sl C1}- and {\sl C2}-optimal designs) are found to be more efficient compared the the other competing criteria. The results show that the modified criteria based optimal designs are more robust under the change of optimality criteria compared to the compound criteria based optimal designs.


 \begin{table}[H]
  \small
  \centering
  \caption{Estimated criterion for  robustness under different optimality criteria  for  optimal designs with $n=16$ runs}
\medskip
  \label{RCC2}
  \begin{tabular}{ccccccc}
\hline
&\multicolumn{6}{c}{Optimality criteria} \\ \cline{2-7} 
Optimal design&{\sl A}&{\sl D}&{\sl DP}&{\sl AP}&{\sl C1}&{\sl C2}\\ \hline 
   $\xi^\star_{\text{\tiny \sl A}, \text{\tiny 2}}$ &        & 0.998 & -- & -- & -- & --  \\ 
  $\xi^\star_{\text{\tiny \sl A}, \text{\tiny 3}}$  &        & 1.000 & 1.000 & 1.000 & 0.829 & 0.803\\ 
   $\xi^\star_{\text{\tiny \sl A}, \text{\tiny 4}}$ &        & 1.000 & -- & -- & -- & -- \\ 
   
\hline
  $\xi^\star_{\text{\tiny \sl D}, \text{\tiny 2}}$  & 0.984 &         & 0.017 & 0.034 & 0.578 & 0.642\\ 
   $\xi^\star_{\text{\tiny \sl D}, \text{\tiny 3}}$ &  1.000 &        & 1.000 & 1.000 & 0.829 & 0.803 \\ 
   $\xi^\star_{\text{\tiny \sl D}, \text{\tiny 4}}$ &   1.000 &       & -- & -- & -- & --  \\
\hline
  $\xi^\star_{\text{\tiny \sl DP}, \text{\tiny 2}}$ &  1.000 & 0.998 &        & 1.000 & 0.872 & 0.839 \\ 
 $\xi^\star_{\text{\tiny \sl DP}, \text{\tiny 3}}$  &   1.000 & 1.000 &       & 1.000 & 0.829 & 0.803  \\ 
  $\xi^\star_{\text{\tiny \sl DP}, \text{\tiny 4}}$ &  0.650 & 0.831 &        & 0.831 & 0.908 & 0.833  \\ \hline
   $\xi^\star_{\text{\tiny \sl AP}, \text{\tiny 2}}$&   1.000 & 0.998 & 1.000 &      & 0.872 & 0.839\\
  $\xi^\star_{\text{\tiny \sl AP}, \text{\tiny 3}}$ &   1.000 & 1.000 & 1.000 &       & 0.829 & 0.803\\
  $\xi^\star_{\text{\tiny \sl AP}, \text{\tiny 4}}$ &  0.863 & 0.930 & 0.962 &         & 0.972 & 0.933\\ \hline
   $\xi^\star_{\text{\tiny \sl C1}, \text{\tiny 2}}$ &    0.991 & 1.000 & 0.628 & 0.719 &      & 0.989  \\
   $\xi^\star_{\text{\tiny \sl C1}, \text{\tiny 3}}$ &   0.857 & 0.866 & 0.503 & 0.591 &     & 1.000 \\
   $\xi^\star_{\text{\tiny \sl C1}, \text{\tiny 4}}$ &   0.895 & 0.951 & 0.782 & 0.889 &     & 0.977\\ \hline
  $\xi^\star_{\text{\tiny \sl C2}, \text{\tiny 2}}$ & 0.960 & 0.987 & 0.552 & 0.726 & 0.995 &       \\
 $\xi^\star_{\text{\tiny \sl C2},  \text{\tiny 3}}$ &  0.857 & 0.866 & 0.503 & 0.591 & 1.000  &       \\
  $\xi^\star_{\text{\tiny \sl C2}, \text{\tiny 4}}$ &    0.960 & 0.987 & 0.552 & 0.726 & 0.995 &       \\ \hline

\end{tabular}
\end{table} 
\label{Saddam}

\section{Conclusion}\label{Conclusion}

In this paper, robustness properties of the  standard, modified and compound criteria based optimal designs are examined by considering regression models for $3^3$ factorial experiment. Three types of robustness criteria, namely robustness under missing observations, under different model assumptions and under the change of the optimality criteria, are considered for comparing robustness properties  of the optimal designs. Robustness under missing observations shows that standard and compound criteria based optimal designs are more  robust compared to the modified criteria based based optimal designs and the compound criteria based designs  are recommended in practice because it correspond to the sufficient number of pure error df  that allows to compare the size of the lack of fit sum of squares. Compound criteria based optimal  designs are found to be the most robust compared to the standard and modified criteria based  optimal designs in terms of robustness under model assumptions. The modified criteria based optimal designs are found to be more robust than the compound criteria based designs when robustness under the change of the optimality criteria is considered and standard criteria based designs are found to be the least robust in this case.


\appendix
\section{Different optimal designs}
\begin{table}[H]
\centering
\small
 \caption{Optimal designs for first degree model $(M_1)$ for three three-level factors in $n=16$ runs.}\medskip
\label{MIS1}
 \begin{tabular}{rrrcccccccc}
\hline
\multicolumn{3}{c}{Design points}&&&\multicolumn{6}{c}{Criterion}\\ \cline{1-3} \cline{6-11} 
 ${x}_{1}$& ${x}_{2}$& ${x}_{3}$ &&&\textit{D} & \textit{A}& \textit{DP} & \textit{AP}&\textit{C1} & \textit{C2} \\ \hline 
 $-1$ & $-1$ & $-1$  &&&  2&   3&   &   &    1&2   \\ 
 $1$  & $-1$ & $-1$  &&&  2&   1&    4&  4&  2&1   \\
$-1$  & $1$  & $-1$  &&&  2&   1&    4&  4&  2&2   \\
 $1$  & $1$  & $-1$  &&&  2&   3&    &   &   1&1   \\ 
$-1$  & $-1$ & $1$   &&&  2&   1&   4&  4&   2&1   \\
 $1$  & $-1$ & $1$   &&&  2&   3&    &   &   1&1   \\ 
$-1$  & $1$  & $1$   &&&  2&   3&    &   &   1&2   \\
 $1$  & $1$  & $1$   &&&  2&   1&   4&  4&   2&1   \\   
 $0$  & $-1$ & $1$   &&&   &    &    &   &   1&1   \\  
 $1$  & $0$  & $-1$   &&&   &    &    &   &   1&1   \\  
 $0$  & $1$  & $1$   &&&   &    &    &   &   1&   \\   
 $0$  & $-1$ & $-1$  &&&   &    &    &   &    &1   \\ 
$-1$  & $0$  & $-1$  &&&   &    &    &   &    1&   \\
$1$  & $0$ & $1$   &&&   &    &    &   &    &1  \\
 $1$  & $1$  & $0$   &&&   &    &    &   &    &1  \\
 
\hline 
\multicolumn{3}{c}{Criterion value}&&&16.00& 16.00&  4.58&  3.37&  8.56&  8.33\\ \hline 
 \end{tabular}
\end{table}

\begin{table}[H]
\centering
 \caption{Optimal designs for second degree polynomial model excluding interaction terms $(M_2)$ for three three-level factors in $n=16$ runs.}\medskip
\label{MIS2}
 \begin{tabular}{rrrcccccccc}
\hline
\multicolumn{3}{c}{Design points}&&&\multicolumn{6}{c}{Criterion}\\ \cline{1-3} \cline{6-11} 
 ${x}_{1}$& ${x}_{2}$& ${x}_{3}$ &&&\textit{D} & \textit{A}& \textit{DP} & \textit{AP}&\textit{C1} & \textit{C2} \\ \hline 
 $-1$ & $-1$ & $-1$  &&&  &   1&    &   2&  &   \\
 $1$  & $-1$ & $-1$  &&&  1&   &    &   &   &1   \\  
$-1$  & $0$  & $-1$  &&&  1&   &    2&  &   &1   \\
 $0$  & $1$  & $-1$  &&&  &   &     &   &   &   \\
 $0$  & $-1$ & $0$   &&&  &   1&    &   &   &   \\ 
$-1$  & $0$  & $0$   &&&  1&   &    &   1&  1&   \\
$-1$  & $1$  & $0$   &&&  1&   1&   2&  &   &1   \\
$-1$  & $-1$ & $1$   &&&  1&   1&   2&  &   2&2   \\ 
 $1$  & $0$  & $1$   &&&  2&   &    2&  2&  &2   \\ 
$-1$  & $1$  & $1$   &&&  &   &     &   2&  &   \\
 $1$  & $1$  & $1$   &&&  &   1&    &   &   1&   \\ 
 $0$  & $0$  & $-1$  &&&  1&   1&   &   2&  &1   \\ 
 $1$  & $0$  & $0$   &&&  &    1&   &   &   &   \\
 $1$  & $1$  & $0$   &&&  &    &    &   &   1&1   \\ 
 $0$  & $1$  & $1$   &&&  1&    1&  2&  &   1&2   \\ 
 $1$  & $1$  & $-1$  &&&   1&   1&  2&  2&  &1   \\
 $1$  & $-1$ & $0$   &&&   1&   &   2&  2&  1&1   \\  
 $0$  & $1$  & $0$   &&&   1&   1&  &   1&  1&   \\  
 $0$  & $-1$ & $1$   &&&   1&   &   &   2&  &   \\
 $0$  & $0$  & $1$   &&&   &   1&   &   &   1&   \\
 $0$  & $0$  & $0$   &&&   1&    &  1&  &   1&1   \\ 
$-1$  & $-1$ & $0$   &&&   &    &   &   &   &   \\
$-1$  & $0$  & $1$   &&&   &   1 &  &   &   &   \\ 
 $0$  & $-1$ & $-1$  &&&   1&   1&  1&  &   2&1   \\ 
$-1$  & $1$  & $-1$  &&&   1&   1&  &   &   2&1   \\   
 $1$  & $-1$ & $1$   &&&   &    1&  &   &   &   \\   
 $1$  & $0$  & $-1$  &&&   &    1&  &   &   2&   \\ 
\hline
\multicolumn{3}{c}{Criterion value}&&&6.00&  7.32&  1.55&  1.31&  7.26&  7.29\\ \hline 
 \end{tabular}
\end{table}

\begin{table}[H]
\centering
 \caption{Optimal designs for second degree polynomial model excluding quadratic terms $(M_3)$ for three three-level factors in $n=16$ runs.}\medskip
\label{MIS3}
 {\begin{tabular}[l]{@{}rrrcccccccc}
\hline
\multicolumn{3}{c}{Design points}&&&\multicolumn{6}{c}{Criterion}\\ \cline{1-3} \cline{6-11} 
 ${x}_{1}$& ${x}_{2}$& ${x}_{3}$ &&&\textit{D} & \textit{A}& \textit{DP} & \textit{AP}&\textit{C1} & \textit{C2} \\ \hline 
 $-1$ & $-1$ & $-1$  &&&   2&	2&   2&  2&  2&  2\\ 
 $1$  & $-1$ & $-1$  &&&   2&   2&   2&  2&  1&  2\\  
$-1$  & $-1$ & $1$   &&&   2&   2&   2&  2&  2&  2\\ 
 $1$  & $-1$ & $1$   &&&   2&   2&   2&  2&  1&  2\\  
$-1$  & $1$  & $-1$  &&&   2&   2&   2&  2&  2&  1\\ 
 $1$  & $1$  & $-1$  &&&   2&   2&   2&  2&  1&  1\\
$-1$  & $1$  & $1$   &&&   2&   2&   2&  2&  2&1  \\
 $1$  & $1$  & $1$   &&&   2&   2&   2&  2&  1&1  \\
 $1$  & $0$  & $-1$  &&&   &    &    &   &   1&  \\ 
 $0$  & $1$  & $-1$  &&&   &	&    &   &   &1   \\ 
 $1$  & $-1$ & $0$   &&&   &    &    &   &   1&  \\ 
$-1$  & $1$  & $0$   &&&   &    &    &   &   & 1  \\ 
 $1$  & $1$  & $0$   &&&   &    &    &   &   1&1  \\ 
 $1$  & $0$  & $1$   &&&   &    &    &   &   1&  \\  
 $0$  & $1$  & $1$   &&&   &    &    &   &   &1   \\  
\hline
\multicolumn{3}{c}{Criterion value}&&&16.00& 16.00&  4.47&  3.01&  8.58&  8.19\\ \hline 
 \end{tabular}}
\end{table}

\begin{table}[H]
\centering
 \caption{Optimal designs for second degree polynomial model $(M_4)$ for three three-level factors in $n=16$ runs.}\medskip
\label{MIS4}
 {\begin{tabular}[l]{@{}rrrcccccccccc}\hline 
\multicolumn{3}{c}{Design points}&&&\multicolumn{6}{c}{Design criterion}\\ 
\cline{1-3}
\cline{6-11}
 ${x}_{1}$& ${x}_{2}$& ${x}_{3}$ &&&{\sl D} & {\sl A}& {\sl DP} & {\sl AP}&{\sl C1} & {\sl C2} \\ \hline
$ 1$&$ 1$&$ 1$&&&1&	  1&1&1 &2		&1\\ 
$-1$&$ 1$&$ 1$&&&1&	  1&2&1	&1		&1\\
$-1$&$ 1$&$-1$&&&1&	  1&&1	&1		&1\\
$ 1$&$ 1$&$-1$&&&1&	  1&&2	&2		&\\
$ 1$&$-1$&$-1$&&&1&	  1&1&1	&1		&1\\
$-1$&$-1$&$-1$&&&1&	  1&2&1	&1		&2\\
$-1$&$-1$&$ 1$&&&1&	  1&1&2	&1		&2\\
$ 1$&$-1$&$ 1$&&&1&	  1&2&1	&1		&2\\
$ 0$&$-1$&$ 0$&&&1&	  &	1&2	&1		&1\\
$ 1$&$ 1$&$ 0$&&&&	  &	&	&		&1\\
$ 0$&$ 1$&$-1$&&&1&	  &	2&  & 		&1\\
$ 0$&$ 0$&$-1$&&&&    &	&	&2		&\\
$ 1$&$ 0$&$ 0$&&&1&   1&2&2 &		&\\
$-1$&$ 1$&$ 0$&&&1&   1&&   &		 &\\
$-1$&$ 0$&$ 1$&&&&    1&&	&		 &\\
$ 0$&$ 1$&$ 1$&&&&    &	&	&		 &\\
$ 1$&$ 0$&$-1$&&&1&	  &	&	&		  &1\\
$-1$&$ 0$&$ 0$&&&&	  &	&	&2		 &1\\
$ 1$&$-1$&$ 0$&&&&	  &	&	&		 &\\
$ 0$&$ 0$&$ 1$&&&1&	  1&2&2	& 		&1\\
$ 1$&$ 0$&$ 1$&&&&	  &	&	&		 &\\
$ 0$&$-1$&$-1$&&& 1&   1&&	&		&\\ 
$-1$&$-1$&$ 0$&&& &    1&&	&		&\\ 
$ 0$&$-1$&$ 1$&&& &    &&	&		&\\ 
$ 0$&$ 1$&$ 0$&&&&1&	 & 	&1		&\\ 
$ -1$&$ 0$&$ -1$&&&1&  1&&	&		&\\ \hline 
\multicolumn{3}{c}{Criterion value}&&&6.72&  7.75&  1.36&  1.01&  7.39&  7.32\\ \hline 
 \end{tabular}}
\end{table}

\end{document}